\documentclass[13 pt]{article}
\usepackage{amssymb}

\def\hcal{{\Large{\textit{h}}}}
\def\dcal{{\Large{\textit{d}}}}
\def\acal{{\Large{\textit{a}}}}
\def\bcal{{\Large{\textit{b}}}}
\def\ccal{{\Large{\textit{c}}}}

\def\Acalt{\tilde{{{\cal A}}}}
\def\Bcalt{\tilde{{{\cal B}}}}
\def\Ccalt{\tilde{{{{\cal C}}}}}

\def\binomNm{\left(^{N}_{m}\right)}
\def\binombN{\left(^{N}_{n+1}\right)}
\def\binomaN{\left(^{N}_{n}\right)}
\def\binomcN{\left(^{N}_{n-1}\right)}
\def\binomNmun{\left(^{N-1}_{n}\right)}

\def\binomNs{\left(^{N}_{s}\right)}
\def\binomNspun{\left(^{N}_{s+1}\right)}
\def\binomNsmun{\left(^{N}_{s-1}\right)}

\newcommand{\beq}{\begin{equation}}
\newcommand{\eeq}{\end{equation}}
\newcommand{\bea}{\begin{eqnarray*}}
\newcommand{\eea}{\end{eqnarray*}}
\newcommand{\beqa}{\begin{eqnarray}}
\newcommand{\eeqa}{\end{eqnarray}}

\begin{document}

\newfont{\elevenmib}{cmmib10 scaled\magstep1}%

\newcommand{\Title}[1]{{\baselineskip=26pt \begin{center}
            \Large   \bf #1 \\ \ \\ \end{center}}}
\hspace*{2.13cm}%
\hspace*{1cm}%
\newcommand{\Author}{\begin{center}\large
           Pascal Baseilhac\footnote{
baseilha@phys.univ-tours.fr} 
\end{center}}
\newcommand{\Address}{{\baselineskip=18pt \begin{center}
\it Laboratoire de Math\'ematiques et Physique Th\'eorique CNRS/UMR 6083,\\
F\'ed\'eration Denis Poisson,\\
Universit\'e de Tours, Parc de Grandmont, 37200 Tours, France\end{center}}}
\baselineskip=13pt

\bigskip
\vspace{-1cm}

\Title{New results in the XXZ open spin chain\\ \vspace{1mm}
{\sffamily \mdseries Proceedings - RAQIS 2007}}\Author

\vspace{- 7mm}
 \Address

\vskip 0.6cm

\centerline{\bf Abstract}\vspace{0.3mm}  \vspace{1mm}
In this review, I describe a recent approach based on the representation theory of the $q-$Onsager algebra which is used to derive exact results for the XXZ open spin chain. 
The complete spectrum and eigenstates are obtained as rational functions of a single variable which discrete values correspond to the roots of a certain characteristic polynomial. Comments and open problems are also presented.
\vspace{0.1cm} 

\vskip 0.8cm

\vskip -0.6cm


\section{Introduction}
Since Sklyanin's seminal work \cite{Sklya} on spin chains with integrable boundary conditions, finding exact results such as the energy spectrum of a model, corresponding eigenstates and correlation functions for elementary excitations has remained an interesting problem in connection with condensed matter or high-energy physics. Among the known integrable open spin chains that have been considered in details, one finds the XXZ open spin chain. For generic boundary conditions, its Hamiltonian reads
\beqa
H_{XXZ}&=&\sum_{k=1}^{N-1}\Big(\sigma_1^{k}\sigma_1^{k+1}+\sigma_2^{k}\sigma_2^{k+1} + \Delta\sigma_3^{k}\sigma_3^{k+1}\Big) \nonumber\\
&&+\ \frac{(q^{1/2}-q^{-1/2})}{(\epsilon^{(0)}_+ + \epsilon^{(0)}_-)}
\Big(  \frac{(\epsilon^{(0)}_+ - \epsilon^{(0)}_-)}{2}\sigma^1_3 + \frac{2}{(q^{1/2}-q^{-1/2})}\big(k_+\sigma^1_+ + k_-\sigma^1_-\big)       \Big)\nonumber\\
 &&+\ \frac{(q^{1/2}-q^{-1/2})}{(\kappa + \kappa^*)}
\Big(  \frac{(\kappa - \kappa^*)}{2}\sigma^N_3 + 2(q^{1/2}+q^{-1/2})\big(\kappa_+\sigma^N_+ + \kappa_-\sigma^N_-\big)       \Big)\label{H}\ ,
\eeqa
where $\sigma_{1,2,3}$ and $\sigma_\pm=(\sigma_1\pm i\sigma_2)/2$ are usual Pauli matrices and $\Delta=(q^{1/2}+q^{-1/2})/2$\ \ denotes the anisotropy parameter. Here, we restrict our attention to the (massless) regime $-1\leq\Delta\leq 1$ i.e. $q=\exp(\phi)$ with $\phi$ purely imaginary. To exhibit the {\it six} independent boundary parameters, below we will sometimes use the following parametrization with $\theta,{\tilde\theta}\in{\mathbb R}$ and ${\alpha}, {\tilde\alpha}\in{\mathbb C}$:
\beqa
&&\epsilon^{(0)}_{+}=(\epsilon^{(0)}_{-})^{\dagger}=\cosh\alpha\ ,\qquad \qquad k_+=(k_{-})^{\dagger}=-(q^{1/2}-q^{-1/2})e^{i\theta}/2 \qquad\ \quad  \ \mbox{(left)}\ ,\nonumber\\ 
&&\kappa^*=(\kappa)^{\dagger}=-\cosh{\tilde\alpha}\ ,\quad \ \qquad \ \ \
\kappa_+=-(\kappa_{-})^{\dagger}=-e^{i{\tilde{\theta}}}/(2(q^{1/2}+q^{-1/2}))\qquad \mbox{(right)}\ .\label{param}
\eeqa

Known to be integrable since 1988, for the simplest {\it diagonal} boundary conditions i.e. $k_\pm=\kappa_\pm\equiv 0$ the algebraic Bethe ansatz approach has been succesfully applied and used to derive the spectrum, eigenstates \cite{Sklya} and more recently correlations functions \cite{Kit} of the model. The derivation of these results essentially relies on the fact that the pseudo-vacuum (or reference) state of this particular case of open spin chain actually coincides with the one associated with the XXZ spin chain with {\it periodic} boundary conditions. 

For non-diagonal boundary conditions, there is no simple or obvious pseudo-vacuum state. From 2001 to 2007, several attempts \cite{Nepo,Cao,WL,Nepo2} were considered in order to circumvent this difficulty: either based on the functional Bethe ansatz approach, or based on the algebraic Bethe ansatz approach using suitable gauge transformations of the double-row monodromy matrix provided the boundary parameters satisfy certain linear relations or $q$ is a root of unity. In any case, either the spectrum was obtained but not the eigenstates or the spectrum and eigenstates were obtained but only for certain sets of boundary conditions. The main advantage of a Bethe-type solution being its efficiency in studying the thermodynamic limit of the spin chain through the link between Bethe equations and non-linear integral equations, it may explain why Bethe ansatz approaches (functional or algebraic) have deserved such attention in all attempts at solving the XXZ open spin chain (\ref{H}). However, for generic boundary parameters and $q$, these standard approaches have failed. Note however a rather simple functional approach recently proposed for generic parameters, where the spectrum is written in terms of solutions of a system of {\it highly transcendental equations} which are not Bethe equations \cite{Galleas}.

Motivated by the challenging problem of solving the XXZ open spin chain (\ref{H}) in the most general regime of integrable boundary conditions, in 2005 we decided to explore a {\it different} and {\it non-standard} (compared to Bethe ansatz) path which takes its roots in Onsager's original article on the solution of the planar Ising model \cite{Ons}. 
A novel approach followed, essentially based on the analysis of the representation theory of the $q-$Onsager algebra. After some efforts \cite{qOns0,qOns,TDpair}, we used this approach to derive the spectrum and the eigenstates of the XXZ open spin chain (\ref{H}) for generic boundary conditions and $q$, as well as we recovered the known results for special relations between left and right boundary parameters \cite{spectXXZ}. 

In this review, our main objective will be to present the novel approach and methodology we have developped in the last two years and describe the main results rather than the technical aspects or details which can be found in our papers \cite{qOns0,qOns,TDpair,spectXXZ} written in collaboration with K. Koizumi. In addition, comments and further directions are briefly presented.

\section{The strategy}
The starting point of the analysis was the discovery that the transfer matrix $t_{XXZ}(u)$ generating (\ref{H}) can be written in terms of a finite set of mutually commuting quantities ${{\cal I}}_{2l+1}$ which form a $q-$deformation of the Onsager's or Dolan-Grady hierarchy \cite{DG,Ahn}. Namely, one has:
\beqa
t_{XXZ}(u)= \sum_{l=0}^{N-1}{\cal F}_{2l+1}(u)\ {{\cal I}}_{2l+1} + {\cal F}_0(u)\ I\!\!I \qquad \mbox{with} \qquad {{\cal I}}_{2l+1}=\kappa {\cal W}^{(N)}_{-l} + \kappa^* {\cal W}^{(N)}_{l+1} + \frac{\kappa_+}{k_+} {\cal G}^{(N)}_{l+1} 
+ \frac{\kappa_-}{k_-} {\tilde{\cal G}}^{(N)}_{l+1}\ \label{tfin}
\eeqa
where ${\cal F}_{2l+1}(u), {\cal F}_{0}(u)$ are rational functions that have been explicitely calculated in \cite{qOns}. In this description, note that the remaining boundary parameters in (\ref{param}) enter in the explicit realizations of the $4N$ fundamental operators \cite{qOns}.
Such remarkable structure comes from the fact that the elements of the double-row monodromy matrix (also called Sklyanin's operator) ${\cal T}(u)$ - $u$ being the spectral parameter - associated with the XXZ open spin chain (\ref{H}) can be decomposed on a basis of $4N$ operators ${\cal W}^{(N)}_{-k},{\cal W}^{(N)}_{k+1},{\cal G}^{(N)}_{k+1},{\tilde
{\cal G}}^{(N)}_{k+1}$, $k=0,1,...,N-1$, formally as (details can be found in \cite{qOns}):  
\beqa {\cal T}(u) =\left(
\begin{array}{cc} 
    {\cal A}(u;\{{\cal W}^{(N)}_{-k},{\cal W}^{(N)}_{k+1}\})  &  {\cal B}(u;\{{\cal G}^{(N)}_{k+1}\})  \\
  {\cal C}(u;\{{\tilde
{\cal G}}^{(N)}_{k+1}\})  & {\cal D}(u;\{{\cal W}^{(N)}_{-k},{\cal W}^{(N)}_{k+1}\})    \\
\end{array} \right) \ . \label{T}\eeqa
In addition, using the fact that the elements of the double-row monodromy matrix satisfy the reflection algebra it is possible to show that these (spectral parameter independent) operators - which act on $N$-tensor product representations of $U_{q^{1/2}}(sl_2)$ - induce a finite dimensional representation of the $q-$deformed analogue of the Onsager infinite dimensional algebra \cite{qOns0} with defining relations\,\footnote{The $q-$commutator \ $[X,Y]_{q}=q^{1/2}XY-q^{-1/2}YX$\ is introduced and the scalar $
\rho = (q^{1/2}+q^{-1/2})^2 k_+k_-$ for the XXZ open spin chain.} for $k,l\in {\mathbb N}$
\beqa
&&\big[{\textsf W}_0,{\textsf W}_{k+1}\big]=\big[{\textsf W}_{-k},{\textsf W}_{1}\big]=\frac{1}{(q^{1/2}+q^{-1/2})}\big({\tilde{\textsf G}_{k+1} } - {{\textsf G}_{k+1}}\big)\ ,\nonumber\\
&&\big[{\textsf W}_0,{\textsf G}_{k+1}\big]_q=\big[{\tilde{\textsf G}}_{k+1},{\textsf W}_{0}\big]_q=\rho{\textsf W}_{-k-1}-\rho{\textsf W}_{k+1}\ ,\nonumber\\
&&\big[{\textsf G}_{k+1},{\textsf W}_{1}\big]_q=\big[{\textsf W}_{1},{\tilde{\textsf G}}_{k+1}\big]_q=\rho{\textsf W}_{k+2}-\rho{\textsf W}_{-k}\ ,\nonumber\\
&&\big[{\textsf W}_0,{\textsf W}_{-k}\big]=0\ ,\quad 
\big[{\textsf W}_1,{\textsf W}_{k+1}\big]=0\ ,\quad \nonumber\\
&&\big[{\textsf G}_{k+1},{\textsf G}_{l+1}\big]=0\ ,\quad   \big[{\tilde{\textsf G}}_{k+1},\tilde{{\textsf G}}_{l+1}\big]=0\ ,\quad
\big[{\tilde{\textsf G}}_{k+1},{\textsf G}_{l+1}\big]
+\big[{{\textsf G}}_{k+1},\tilde{{\textsf G}}_{l+1}\big]=0\ .\label{qOns}
\eeqa

In light of the above formulation, it was tempting to consider the XXZ open spin chain (\ref{H}) from the point of view of the representation theory of the algebra (\ref{qOns}) in view of the success of the Onsager's algebra and related representation theory in solving the Ising model \cite{Ons}, generalizations \cite{Ahn} and superintegrable chiral Potts model \cite{Potts,Davies}. Having this in mind, we implemented the following procedure in order to solve the spectral problem for (\ref{H}) or, more generally, the transfer matrix:\vspace{2mm}

$\bullet$ Use the relation between the algebra (\ref{qOns}) and Terwilliger's tridiagonal algebra \cite{Ter}. See \cite{qOns0};\vspace{1mm}

$\bullet$ Show that ${\cal W}^{(N)}_{0},{\cal W}^{(N)}_{1}$ form a tridiagonal pair (see below). See \cite{qDG,qOns0,TDpair};\vspace{1mm}

$\bullet$ Construct the states $\{\psi^{(N)}_{n[j]}\}$ and $\{\varphi^{(N)}_{s[k]}\}$ on which ${\cal W}^{(N)}_{0},{\cal W}^{(N)}_{1}$ act as block tridiagonal matrices. See \cite{TDpair};\vspace{1mm}

$\bullet$ Show that ${\cal W}^{(N)}_{-k},{\cal W}^{(N)}_{k+1},{\cal G}^{(N)}_{k+1},{\tilde
{\cal G}}^{(N)}_{k+1}$, $k=1,...,N-1$ also act as block tridiagonal matrices; See \cite{spectXXZ}.\vspace{1mm}

$\bullet$ Write the spectral problem for all quantities ${\cal I}_{2l+1}$ as:

\qquad \qquad $\rightarrow$ a {\it coupled system three-term recursion relations};

\qquad \qquad $\rightarrow$ a {\it system of second-order $q-$difference equations};\vspace{1mm}

\quad and derive the spectrum and eigenstates. See \cite{spectXXZ}.

\vspace{2mm}
 
It is important to stress that the above procedure introduced to derive all eigenvalues and eigenstates of (\ref{tfin}) solely uses the {\it block tridiagonal} structure of the commuting operators ${\cal I}_{2l+1}$ in two suitable basis. However, similarly to the Ising or superintegrable model \cite{Davies}, it is possible to show that the operators ${\cal W}^{(N)}_{-k},{\cal W}^{(N)}_{k+1},{\cal G}^{(N)}_{k+1},{\tilde
{\cal G}}^{(N)}_{k+1}$, $k=0,...,N$ satisfy a set of linear relations. Exploiting this property should lead to {\it another} solution - may be simpler - to the spectral problem for the transfer matrix of the XXZ open spin chain (\ref{H}). We will explain this point in the last Section.

\section{Solving the spectral problem} 
\subsection{Mathematical background}
Almost nothing being known about the representation theory of the algebra (\ref{qOns}), we used its connection with Terwilliger's tridiagonal algebra \cite{Ter} (sometimes called the $q-$Onsager algebra). Indeed, we had previously noticed \cite{qOns} (see also \cite{qDG}) that ${\cal W}^{(N)}_{0},{\cal W}^{(N)}_{1}$ given by
\beqa
{\cal W}^{(N)}_0&=& (k_+\sigma_+ + k_-\sigma_-)\otimes I\!\!I^{(N-1)} + q^{\sigma_3/2}\otimes {\cal W}_0^{(N-1)}, \nonumber \\
{\cal W}^{(N)}_1&=& (k_+\sigma_+ + k_-\sigma_-)\otimes I\!\!I^{(N-1)} + q^{-\sigma_3/2}\otimes {\cal W}_1^{(N-1)}\label{op}
\eeqa
with ${\cal W}^{(0)}_0\equiv \epsilon_+^{(0)}$\ ,  ${\cal W}^{(0)}_1\equiv \epsilon_-^{(0)}$\ 
satisfy the so-called $q-$Dolan-Grady relations
\beqa \big[{\cal W}^{(N)}_0,\big[{\cal W}^{(N)}_0,\big[{\cal W}^{(N)}_0,{\cal
W}^{(N)}_1\big]_q\big]_{q^{-1}}\big]&=&\rho\big[{\cal W}^{(N)}_0,{\cal W}^{(N)}_1\big]\ ,\nonumber\\
\big[{\cal W}^{(N)}_1,\big[{\cal W}^{(N)}_1,\big[{\cal W}^{(N)}_1,{\cal
W}^{(N)}_0\big]_q\big]_{q^{-1}}\big]&=&\rho\big[{\cal W}^{(N)}_1,{\cal W}^{(N)}_0\big]\ 
\label{Talg}
\eeqa
and induce a representation of the tridiagonal algebra. According to Terwilliger's work \cite{Ter}, it follows that ${\cal W}^{(N)}_{0},{\cal W}^{(N)}_{1}$ provide an explicit example of  {\it tridiagonal pair}. Roughly speaking, for generic boundary parameters (\ref{param}) and generic  values of $q$ it means that there exists a complete basis (resp. `dual' basis) of ${\cal V}$ in which  ${\cal W}_{0}^{(N)}$ (resp. ${\cal W}_{1}^{(N)}$) is represented by a diagonal matrix with degeneracies and ${\cal W}_{1}^{(N)}$ (resp. ${\cal W}_{0}^{(N)}$) is represented by a block tridiagonal matrix. Such basis and dual one have been constructed in \cite{TDpair}, and are denoted below $\{\psi^{(N)}_{n[j]}\}$ and $\{\varphi^{(N)}_{s[k]}\}$. 

Let us recall the main results of \cite{TDpair}. Using the canonical basis $\bigotimes_{j=1}^{N}|\pm\rangle_{j}$ on which the nonlocal operators (\ref{op}) act, let us introduce the $2^N$ ordered states 
\beqa
\psi^{(N)}_{n[j]}&=&\Big( e^{\alpha+(N-1-2n)\phi/2+i\theta}|+\rangle_{N} + |-\rangle_{N} \Big)\otimes \psi^{(N-1)}_{n[j]}\ \quad \quad\ \ \quad \quad\quad \mbox{for}\quad j\in \{1,...,\Big({{N-1}\atop {n}}\Big)\}\ ,\nonumber\\
\psi^{(N)}_{n[j]}&=&\Big(e^{-\alpha-(N+1-2n)\phi/2+i\theta}|+\rangle_{N} + |-\rangle_{N} \Big)\otimes \psi^{(N-1)}_{n-1[j-\binomNmun]}\ \quad \quad \ \mbox{for}\quad j\in \{\Big({{N-1}\atop {n}}\Big)+1,...,\Big({{N}\atop {n}}\Big)\}\ .\label{eigenvectN}
\eeqa
For generic boundary parameters and $q$, these states form a complete\,\footnote{For special relations among the boundary parameters, the representation may become indecomposable. We do not consider such possibility here.} basis of ${\cal V}$ on which the tridiagonal pair ${\cal W}_{0}^{(N)}$, ${\cal W}_{1}^{(N)}$ acts \cite{TDpair}. Using the explicit expressions (\ref{op}) it is straightforward to exhibit the diagonal and block tridiagonal structure of the operators ${\cal W}_{0}^{(N)}$, ${\cal W}_{1}^{(N)}$, respectively, in this basis: 
\beqa
{\cal W}^{(N)}_0\psi^{(N)}_{n[j]}&=&\lambda^{(N)}_n\psi^{(N)}_{n[j]} \qquad \mbox{with} \qquad \lambda^{(N)}_n=\cosh(\alpha+(N-2n)\phi/2)\ ,\nonumber\\
{\cal W}^{(N)}_1\psi^{(N)}_{n[j]}&=& \sum_{i=1}^{\binombN}\bcal^{(N,0)}_{n[ij]}\psi^{(N)}_{n+1[i]} + \sum_{i=1}^{\binomaN}\acal^{(N,0)}_{n[ij]}\psi^{(N)}_{n[i]} + \sum_{i=1}^{\binomcN}\ccal^{(N,0)}_{n[ij]}\psi^{(N)}_{n-1[i]} \ \label{recW1}
\eeqa
with $n=0,1,...,N$ and $j\in\{1,...,\big({{ N }\atop {n}}\big)\}$. Note that the explicit form of the coefficients $\acal^{(N,0)}_{n[ij]},\bcal^{(N,0)}_{n[ij]},\ccal^{(N,0)}_{n[ij]}$ can be found in \cite{TDpair}. Clearly, a similar analysis can also be performed by considering the eigenbasis of ${\cal W}^{(N)}_{1}$ instead of ${\cal W}^{(N)}_{0}$, leading to analogous results. In this case, the `dual' states given by ${\varphi}^{(N)}_{s[k]}=\psi^{(N)}_{s[k]}|_{\alpha\rightarrow -\alpha^*,\ \phi\rightarrow - \phi,\ \theta\rightarrow \theta+i\pi}$ form an eigenbasis of ${\cal W}^{(N)}_{1}$ \cite{TDpair}.\vspace{1mm}

All higher operators realizing (\ref{qOns}) being generated recursively from the lower ones \cite{qOns}, there were good reasons to expect that they would act in a simple manner on the eigenbasis of ${\cal W}_{0}^{(N)}$ or ${\cal W}_{1}^{(N)}$. Indeed, proceeding similarly for higher values of $l=1,2,...,N-1$ it is possible to show that  {\it all} nonlocal operators ${\cal W}^{(N)}_{-k},{\cal W}^{(N)}_{k+1},{\cal G}^{(N)}_{k+1},{\tilde{\cal G}}^{(N)}_{k+1}$ {\it also} enjoy a block tridiagonal structure in the basis (\ref{eigenvectN}). After some straightforward calculations, one finds:
\beqa
{\cal W}^{(N)}_{-l}\psi^{(N)}_{n[j]}&=& \sum_{i=1}^{\binomaN}\dcal^{(N,l)}_{n[ij]}\psi^{(N)}_{n[i]}\ ,\nonumber\\ 
{\cal W}^{(N)}_{l+1}\psi^{(N)}_{n[j]}&=& \sum_{i=1}^{\binombN}\bcal^{(N,l)}_{n[ij]}\psi^{(N)}_{n+1[i]} + \sum_{i=1}^{\binomaN}\acal^{(N,l)}_{n[ij]}\psi^{(N)}_{n[i]} + \sum_{i=1}^{\binomcN}\ccal^{(N,l)}_{n[ij]}\psi^{(N)}_{n-1[i]}\ ,\nonumber\\
{\cal G}^{(N)}_{l+1}\psi^{(N)}_{n[j]}&=&\ \ \sum_{i=1}^{\binombN}(q^{1/2}\lambda^{(N)}_{n}-q^{-1/2}\lambda^{(N)}_{n+1})\bcal^{(N,l)}_{n[ij]}\psi^{(N)}_{n+1[i]} + \sum_{i=1}^{\binomaN}\hcal^{(N,l)}_{n[ij]}\psi^{(N)}_{n[i]} \nonumber \\ 
&&+\ \sum_{i=1}^{\binomcN}(q^{1/2}\lambda^{(N)}_{n}-q^{-1/2}\lambda^{(N)}_{n-1})\ccal^{(N,l)}_{n[ij]}\psi^{(N)}_{n-1[i]}\ ,\nonumber\\
{\tilde{\cal G}}^{(N)}_{l+1}\psi^{(N)}_{n[j]}&=&\ \ \sum_{i=1}^{\binombN}(q^{1/2}\lambda^{(N)}_{n+1}-q^{-1/2}\lambda^{(N)}_{n})\bcal^{(N,l)}_{n[ij]}\psi^{(N)}_{n+1[i]} + \sum_{i=1}^{\binomaN}\hcal^{(N,l)}_{n[ij]}\psi^{(N)}_{n[i]} \nonumber \\ 
&&+\ \sum_{i=1}^{\binomcN}(q^{1/2}\lambda^{(N)}_{n-1}-q^{-1/2}\lambda^{(N)}_{n})\ccal^{(N,l)}_{n[ij]}\psi^{(N)}_{n-1[i]}\label{recgen}
\eeqa
where all coefficients can be found in \cite{spectXXZ}. Note that they have a rather simple form regarding to the recursion on the index $l$, a fact which may have been anticipated from the existence of linear relations among the operators. See the comments below. In the dual basis $\{{\varphi}^{(N)}_{s[k]}\}$, similar results are also obtained.

Having two different basis $\{{\psi}^{(N)}_{n[j]}\},\{{\varphi}^{(N)}_{s[k]}\}$ on which all mutually commuting operators generating the $q-$Dolan-Grady hierarchy (\ref{tfin}) act as block tridiagonal matrices with entries known explicitely, we are now ready to consider the spectral problem for the XXZ open spin chain (\ref{H}) formulated using (\ref{tfin}).

\subsection{Eigenstates and spectrum for generic parameters}
For generic values of the boundary parameters and $q$, there is no nontrivial subspace of $\cal V$ which is left invariant under the action of the nonlocal commuting operators ${\cal I}_{2l+1}$. Considering the most general linear combinations of states $\psi^{(N)}_{n[j]}$ or $\varphi^{(N)}_{s[k]}$, it follows that any eigenstate admits two dual representations with respect to the two dual basis on which the elements of the $q-$Onsager algebra act in a block tridiagonal form. Any eigenstate can be written either
\beqa
\Psi(\Lambda_1)= \sum_{n=0}^{N}\sum_{j=1}^{\big({{N}\atop {n}}\big)} f^{(+)}_{n[j]}\big(\Lambda_1\big)\psi^{(N)}_{n[j]}\quad \qquad \mbox{or}\quad 
 \qquad \Psi(\Lambda_1)= \sum_{s=0}^{N}\sum_{k=1}^{\big({{N}\atop {s}}\big)} f^{(-)}_{s[k]}\big(\Lambda_1\big)\varphi^{(N)}_{s[k]}\ ,  \label{vectdualN}
\eeqa
where each of the dual families of weights $\{f^{(+)}_{n[j]}\},\{f^{(-)}_{s[k]}\}$ and the restricted set of `allowed' eigenvalues $\{\Lambda_{1,1},...,\Lambda_{1,2^N}\}$ have to be determined. For generic boundary parameters, the spectrum of ${\cal I}_{1}$ is not degenerate. As $[{\cal I}_{2l+1},{\cal I}_{2k+1}]=0$, it follows that $\{f^{(+)}_{n[j]}\},\{f^{(-)}_{s[k]}\}$ can be uniquely derived from the constraint:
\beqa
{\cal I}_{1}|\Psi(\Lambda_1)\rangle=\Lambda_1|\Psi(\Lambda_1)\rangle\ .
\eeqa
For any representation (\ref{vectdualN}) choosen, this equation leads to a system of {\it coupled three-term recurrence relations} for the weights which, in some sense, generalizes the Askey-Wilson recursion relations\,\footnote{For $N=1$, the coefficients are special cases of the Askey-Wilson ones.} for $q-$orthogonal polynomials. For $n,s=0,1,...,N$ one has:
\beqa
\sum_{m=1}^{\big({{N}\atop {n-1}}\big)}{\cal B}^{(N,0)}_{n-1[jm]}f^{(N,+)}_{n-1[m]}\big(\Lambda_1\big) +\sum_{m=1}^{\big({{N}\atop {n+1}}\big)}{\cal C}^{(N,0)}_{n+1[jm]}f^{(N,+)}_{n+1[m]}\big(\Lambda_1\big)
+ \sum_{m=1}^{\big({{N}\atop {n}}\big)}\big({\cal A}^{(N,0)}_{n[jm]} - \Lambda_1\delta_{jm}\big)f^{(N,+)}_{n[m]}\big(\Lambda_1\big) =0\  , \label{recfM}
\eeqa
or, alternatively, 
\beqa
\sum_{m=1}^{\big({{N}\atop {s-1}}\big)}{\cal B}^{'(N,0)}_{s-1[km]}f^{(N,-)}_{s-1[m]}\big(\Lambda_1\big) +\sum_{m=1}^{\big({{N}\atop {s+1}}\big)}{\cal C}^{'(N,0)}_{s+1[km]}{ f}^{(N,-)}_{s+1[m]}\big(\Lambda_1\big)
+ \sum_{m=1}^{\big({{N}\atop {s}}\big)}\big({\cal A}^{'(N,0)}_{s[km]} - \Lambda_1\delta_{km}\big){ f}^{(N,-)}_{s[m]}\big(\Lambda_1\big) =0\   \label{recfMdual}
\eeqa
where the coefficients ${\cal X}^{(N,0)}_{n[jm]},{\cal X}^{'(N,0)}_{s[km]}$ (${\cal X}\in\{{\cal A},{\cal B},{\cal C}\}$) can be easily obtained from (\ref{recgen}) \cite{spectXXZ}. Having $2^N$ equations in total, one finds that these weights are rational functions of the variable $\Lambda_1\in\{\Lambda_{1,1},...,\Lambda_{1,2^N}\}$ where $\Lambda_{1,r}$ are the roots of the characteristic polynomial of degree $d=2^N$
\beqa
{\cal P}\big(\Lambda_1\big)=\det[{\cal I}_{1}-\Lambda_1I\!\!I]\ .\label{P}\ 
\eeqa
The complete family of eigenstates (\ref{vectdualN}) being identified, the spectrum of the transfer matrix can now be derived.
For instance, acting with higher operators ${\cal I}_{2l+1}$ of the $q-$Dolan-Grady hierarchy on these eigenstates, and taking the scalar product with 
\beqa
{\tilde\varphi}^{(N)}_{s[k]}=\psi^{(N)}_{s[k]}|_{\alpha\rightarrow -\alpha} \qquad \mbox{or} \qquad {\tilde\psi}^{(N)}_{n[j]}={\varphi}^{(N)}_{n[j]}|_{\alpha^*\rightarrow -\alpha^*} \ ,\nonumber
\eeqa
respectively, according to the representation choosen in (\ref{vectdualN}) one ends up with  a {\it second-order $q-$difference equation} that determines all higher necessary eigenvalues  $\Lambda_{2l+1}$, $l=1,...,N-1$ as rational functions of $\Lambda_1$. For any choice of $s,k,n,j$, one writes either
\beqa
\Lambda_{2l+1}&=&   \sum_{m=1}^{\binomNspun}\Bcalt^{(N,l)}_{s[mk]}\frac{{\Psi}^{(N,+)}_{[m]}(s+1)}{{\Psi}^{(N,+)}_{[k]}(s)} + \sum_{m=1}^{\binomNs}\Acalt^{(N,l)}_{s[mk]}\frac{{\Psi}^{(N,+)}_{[m]}(s)}{{\Psi}^{(N,+)}_{[k]}(s)} + \sum_{m=1}^{\binomNsmun}\Ccalt^{(N,l)}_{s[mk]}\frac{{\Psi}^{(N,+)}_{[m]}(s-1)}{{\Psi}^{(N,+)}_{[k]}(s)}\ \ , \label{spectrumIp}\\
\mbox{or} \qquad \Lambda_{2l+1}&=&   \sum_{m=1}^{\binombN}\Bcalt^{'(N,l)}_{n[mj]}\frac{{\Psi}^{(N,-)}_{[m]}(n+1)}{{\Psi}^{(N,-)}_{[j]}(n)} + \sum_{m=1}^{\binomaN}\Acalt^{'(N,l)}_{n[mj]}\frac{{\Psi}^{(N,-)}_{[m]}(n)}{{\Psi}^{(N,-)}_{[j]}(n)} + \sum_{m=1}^{\binomcN}\Ccalt^{'(N,l)}_{n[mj]}\frac{{\Psi}^{(N,-)}_{[m]}(n-1)}{{\Psi}^{(N,-)}_{[j]}(n)}\ ,
\label{spectrumIm}
\eeqa
where the coefficients can be found in \cite{spectXXZ} and the functions
\beqa
{\Psi}^{(N,+)}_{[k]}(s)=  \big({\tilde\varphi}^{(N)}_{s[k]},  \Psi(\Lambda_1)\big)  \qquad \mbox{and} \qquad  {\Psi}^{(N,-)}_{[j]}(n)=  \big({\tilde\psi}^{(N)}_{n[j]},  \Psi(\Lambda_1)\big) \   \label{eigenf2}\nonumber
\eeqa
have been introduced. To resume, all eigenvalues $\Lambda_{2l+1}$, $l=1,...,N-1$ are obtained as rational functions of the single variable $\Lambda_1$ restricted on the discrete support defined by ${\cal P}\big(\Lambda_1\big)=0$ with (\ref{P}).
In terms of these, the spectrum of the Hamiltonian (\ref{H}) reads:
\beqa
E&=& \frac{(q^{1/2}-q^{-1/2})(q^{1/2}+q^{-1/2})^{-1}}{2(\kappa+\kappa^*)(\epsilon_+^{(0)}+\epsilon_-^{(0)})}\left(\sum_{l=0}^{N-1}\frac{d{\cal F}_{2l+1}(u)}{du}|_{u=1}\ \Lambda_{2l+1} + \frac{d{\cal F}_0(u)}{du}|_{u=1}\right)\ \nonumber \\
&&- \left(N\Delta + \frac{(q^{1/2}-q^{-1/2})^2}{2(q^{1/2}+q^{-1/2})}\right)\ .  \label{E}\nonumber
\eeqa

\subsection{Truncation and the Bethe ansatz regime of boundary parameters}
For certain special relations between left and right boundary parameters, important simplifications occur. These situations arise when some of the off-diagonal {\it lower} (resp. {\it upper}) blocks of the block tridiagonal matrices representing ${\cal I}_{2l+1}$ for any $l=0,1,...,N-1$ identically vanish, leading respectively to two different conditions on the boundary parameters. For all $l,i,j,m,k$ and integer $p$ fixed:
\beqa
{\cal B}^{(N,l)}_{p[ij]}\equiv 0 \quad \mbox{and} \quad {\cal B}^{'(N,l)}_{N-p-1[mk]}\equiv 0\  \ \Rightarrow \ \ \alpha\pm{\tilde\alpha} &=& - i({\tilde \theta}-\theta) - (N-2p-1)\phi/2\qquad mod(2i\pi)\ ; \label{r1}\\
{\cal C}^{(N,l)}_{p+1[ij]}\equiv 0 \quad \mbox{and} \qquad {\cal C}^{'(N,l)}_{N-p[mk]}\equiv 0\  \ \Rightarrow \ \ \alpha\pm{\tilde\alpha} &=& \ \ i({\tilde \theta}-\theta) - (N-2p-1)\phi/2 \qquad  mod(2i\pi)\ .\label{r2}
\eeqa
In these two special cases, the coupled three-term recursion relations (\ref{recfM}) or (\ref{recfMdual}) determining the weights in the expansions (\ref{vectdualN}) become truncated. This means that any eigenstate can be written as a truncated combination of one of the two representations in (\ref{vectdualN}). For the first family of boundary conditions (\ref{r1}), the eigenstates read:
\beqa
\Psi^{(+)}(\Lambda_1)= \sum_{n=0}^{p}\sum_{j=1}^{\big({{N}\atop {n}}\big)} f^{(N,+)}_{n[j]}\big(\Lambda_1\big)\psi^{(N)}_{n[j]}\ \qquad \mbox{and}\ \qquad {\Psi}^{(-)}(\Lambda_1)= 
\sum_{s=0}^{N-p-1}\sum_{k=1}^{\big({{N}\atop {s}}\big)} f^{(N,-)}_{s[k]}\big(\Lambda_1\big)
\varphi^{(N)}_{s[k]}\ \label{vecttrunc1}
\eeqa
whereas 
\beqa
\Psi^{(+)}(\Lambda_1)= \sum_{n=p+1}^{N}\sum_{j=1}^{\big({{N}\atop {n}}\big)} f^{(N,+)}_{n[j]}\big(\Lambda_1\big)\psi^{(N)}_{n[j]}\ \qquad \mbox{and}\ \qquad {\Psi}^{(-)}(\Lambda_1)= 
\sum_{s=N-p}^{N}\sum_{k=1}^{\big({{N}\atop {s}}\big)} f^{(N,-)}_{s[k]}\big(\Lambda_1\big)
\varphi^{(N)}_{s[k]}\ \label{vecttrunc1}
\eeqa
for the second family of boundary conditions (\ref{r2}). In others words, the complete family of eigenvalues and eigenstates splits in two sets, each associated with a characteristic polynomial of degree $d_1<2^N$ and $d_2< 2^N$, respectively, such that $d_1+d_2=2^N$. According to the family of boundary conditions
one finds:
\beqa
d_1=\sum_{n=0}^{p}\biggl({{N}\atop {n}}\biggr)\quad \mbox{for} \quad (\ref{r1}) \qquad  \mbox{and} \qquad  d_1=\sum_{n=p+1}^{N}\biggl({{N}\atop {n}}\biggr)\quad \mbox{for} \quad (\ref{r2})\ . \nonumber
\eeqa
Then, the eigenvalues $\Lambda_{2l+1}$ for $l=1,...,N$ are obtained with the same method as for generic boundary parameters. 

It is important to mention that the linear relations (\ref{r1}), (\ref{r2}) among the boundary parameters coincide exactly with the ones for which a Bethe-type solution exists, although the derivation of these relations is here totally different. Also, the splitting of the eigenvalues and eigenstates in two sectors is also confirmed: in the algebraic Bethe ansatz framework \cite{Cao,WL} (see also \cite{Nepo}), the two sets of eigenstates and eigenvalues are described by two different sets of Bethe equations. This shows that the approach described here provides an alternative solution compared to the Bethe ansatz one for the model (\ref{H}) with the family of boundary conditions (\ref{r1}), (\ref{r2}). For these regimes of boundary parameters, a detailed comparison between both approaches is an interesting problem: it should give explicit relations between the solutions of Bethe equations and the roots of the truncated characteristic polynomials.

\section{Comments and open problems}

\qquad $\bullet$ {\bf Is the model `{\it really}' solved?} If one {\it believes} that solving an integrable model {\it only} means that one must be able to find its spectrum\,\footnote{I do not even speak about the eigenstates which can not be derived using the functional Bethe ansatz approach.} in terms of solutions of {\it highly transcendental equations} (Bethe-type equations), then the solution we proposed for the XXZ open spin chain (\ref{H}) \cite{spectXXZ} is indeed not satisfying at all. And from this point of view, this is not the only integrable model which is out of range\,\footnote{Consider for instance the XYZ chain, quantum Toda chain, chiral Potts model,...}! On the other hand, if one thinks that solving an integrable model means that one is able to reduce the complexity of the system using unified mathematical structures (quantum algebra, representation theory) in order to write its spectrum in terms of solutions of simpler equations (for instance the {\it algebraic} equations with (\ref{recfM}), (\ref{recfMdual}) where ${\cal I}_{1}$ is a {\it block tridiagonal} matrix) then from this point of view the model (\ref{H}) is solved. \vspace{2mm} 

$\bullet$ {\bf Large $N$ and thermodynamic limit?} For generic boundary parameters, a natural question to ask is weither diagonalizing ${\cal I}_{1}$ is simpler and faster than diagonalizing ${\cal H}$ and all higher conserved quantities generated from the transfer matrix. Indeed, one has to remember that the eigenstates and eigenvalues of ${\cal I}_{1}$ generate the eigenvalues  of all higher quantities  ${\cal I}_{2l+1}$ for $l=1,...,N-1$. Thanks to the block tridiagonal structure of the matrix ${\cal I}_{1}$, finding the spectrum has been reduced to (\ref{recfM}) or (\ref{recfMdual}). Although this simplified formulation helps for numerical calculations (using well-known FORTRAN procedures for block tridiagonal matrices), for $N>>1$ or in the thermodynamic limit $N\rightarrow\infty$ solving {\it directly} these $2^N$ coupled three-term recurrence relations is a problem that clearly needs further investigations. Studying infinite dimensional representations of (\ref{qOns}) might be a possible direction.  \vspace{2mm} 

$\bullet$ {\bf Another solution using the $q-$Onsager algebra?} The solution here described \cite{spectXXZ} is essentially based on the block tridiagonal structure of all operators satisfying (\ref{qOns}) in a certain basis. However, for finite dimensional representations associated with the XXZ open spin chain (\ref{H}) it is also interesting to point out that\,\footnote{
$C_{-n}^{(N)}=(-1)^{N-n} (q^{1/2}+q^{-1/2})^{n+1}\Big(\frac{2(q+q^{-1})}{(q^{1/2}+q^{-1/2})}\Big)^{N-n-2}
 \frac{(N-1)!}{(n+1)!(N-n-2)!} \left(\frac{2(q+q^{-1})}{(q^{1/2}+q^{-1/2})} \frac{N}{N-n-1} + \frac{\epsilon^{(0)}_{+}\epsilon^{(0)}_{-}(q^{1/2}-q^{-1/2})^2}{k_+k_-(q^{1/2}+q^{-1/2})}\right)\ $.}
\beqa
&&-\frac{(q^{1/2}-q^{-1/2})}{k_+k_-}\omega_0^{(N)}{\cal W}_{0}^{(N)}+\sum^{N}_{l=1}C_{-l+1}^{(N)}{\cal W}_{-l}^{(N)} + \epsilon^{(N)}_{+}I\!\!I^{(N)}=0\ ,\nonumber\\
&&-\frac{(q^{1/2}-q^{-1/2})}{k_+k_-}\omega_0^{(N)}{\cal W}_{1}^{(N)}+\sum^{N}_{l=1}C_{-l+1}^{(N)}{\cal W}_{l+1}^{(N)} + \epsilon^{(N)}_{-}I\!\!I^{(N)}=0\ ,\nonumber\\
&&-\frac{(q^{1/2}-q^{-1/2})}{k_+k_-}\omega_0^{(N)}{\cal G}_{1}^{(N)}+\sum^{N}_{l=1}C_{-l+1}^{(N)}{\cal G}_{l+1}^{(N)}=0\ ,\nonumber\\
&&-\frac{(q^{1/2}-q^{-1/2})}{k_+k_-}\omega_0^{(N)}{\tilde {\cal G}}_{1}^{(N)}+\sum^{N}_{l=1}C_{-l+1}^{(N)}{\tilde {\cal G}}_{l+1}^{(N)}=0\ \label{c4}\ .
\eeqa
For the reader who is familiar with the Ising and superintegrable chiral Potts models, the spectrum
of these models is known to be expressed in terms of the roots of the polynomial (called Baxter's polynomial for the superintegrable Potts model) associated with $\sum_l\alpha_l A_l=0$ and $\sum_l\alpha_l G_l=0$ where $A_l,G_l$ are Onsager's algebra generators (see references and details in \cite{Davies}). By analogy, linear relations of the form (\ref{c4}) may be key ingredients in finding another solution - may be simpler than the one we proposed in \cite{spectXXZ} - to the XXZ open spin chain (\ref{H}).\vspace{2mm}
  
$\bullet$ {\bf Beyond.} The computation of correlation functions essentially relies on the knowledge of the complete set of eigenstates (in the {\it massless} regime which is here considered). In view of (\ref{vectdualN}), one can easily calculate the action of the elementary local operators on the basis (\ref{eigenvectN})
\beqa
\sigma^{(k)}_a \psi^{(N)}_{n[i]} =\sum_{m=0}^N\sum_{j=1}^{\binomNm} {\cal M}^{(N,k)}_{a\ n[i], m[j]} \ \psi^{(N)}_{m[j]}\qquad \mbox{with}\qquad \sigma^{(k)}_a = I\!\!I \otimes ...\otimes I\!\!I \otimes \underbrace{\sigma_a}_{\mbox{\small at site} \ k} \otimes I\!\!I \otimes ...\otimes I\!\!I\nonumber,\quad a\in\{1,2,3\},
\eeqa
and the coefficients ${\cal M}^{(N,k)}_{a\ n[i], m[j]}$ are derived recursively. As an example, it is possible to write one-point functions as rational functions of the variable $\Lambda_1$ defined on the discret support ${\cal P}(\Lambda_1)=0$ with (\ref{P}). For small values of $N<6$ and generic parameters or the (Bethe ansatz) regime of parameters\,\footnote{The calculations are much simpler and faster due to the truncation of the eigenstates.} (\ref{r1}), (\ref{r2}), the results are easily checked numerically. Having a better understanding of solutions to (\ref{recfM}), (\ref{recfMdual}) at large $N$ would then play an essential role for studying correlation functions. In this direction, extending the results of \cite{KuznetsovSkly} might be helpful.

\vspace{0.2cm}

\noindent{\bf Acknowledgements:} I thank the organizers of RAQIS Meeting 2007 where these results have been presented. Part of this work is supported by the ANR research project ``{\it Boundary integrable models: algebraic structures and correlation functions}'', contract number JC05-52749.

\end{document}